\documentclass[12pt]{article}
\usepackage[cp1250]{inputenc}
\usepackage{amsmath}

\usepackage{graphicx}

\renewcommand{\(}{\left(}
\renewcommand{\)}{\right)}
\newcommand{\E}{\mathrm{E}}
\newcommand{\Var}{\mathrm{Var}}

\author{Kamil Jod\'z}
\title{Mortality in a heterogeneous population - Lee-Carter's methodology} 

\begin{document}
	
	\maketitle
	
	\begin{abstract}
		The EU Solvency II directive recommends insurance companies to pay more attention to the risk management methods. The sense of risk management is the ability to quantify risk and apply methods that  reduce uncertainty. In life insurance, the risk is a consequence of the random variable describing the life expectancy. The article will present a proposal for stochastic mortality modeling based on the Lee and Carter methodology. The maximum likelihood method is often used to estimate parameters in mortality models. This method assumes that the population is homogeneous and the number of deaths has the Poisson distribution. The aim of this article is to change assumptions about the distribution of the number of deaths. The results indicate that the model can get a better match to historical data, when the number of deaths has a negative binomial distribution.

		keywords: Mortality rates, Lee-Carter model, heterogeneous population
	
	\end{abstract}

	\section{Introduction}
		In many highly developed countries, the average life expectancy is clearly increasing. Advances in medicine and improving living conditions with each year mean that the average life expectancy has never been as high as it is today. 20-30 years ago, the population aged 65+ was a small percentage of the general population. At the moment, in some countries, the percentage of people aged 65+ is over 20\%. This improvement is a serious challenge for insurance companies and national pension insurance systems.The problem has been noticed by the institutions of the European Union, which results is the Solvency II Directive. This document also refers to the methods of risk management used by insurance companies. In the case of companies offering life insurance, one of the most important factors increasing the risk of insolvency is the random nature of the variable describing life expectancy. Searching for methods that allow for more accurate modeling and better life expectancy forecasting is due to the need to minimize the risk of this stochastic factor. The article structure is as follows: in the second chapter mortality models based on the Lee-Carter methodology are briefly discussed. In all models, it was assumed that the number of deaths has the Poisson distribution, while the maximum likelihood methods were used to estimate the parameters.The third chapter presents the idea and justification for the use of the negative binomial distribution for modeling the number of deaths instead of the Poisson distribution. The fourth chapter contains the results of research conducted on data on the mortality in the Polish population. A comparison of mortality models with different assumptions about the distribution of deaths is presented. The last chapter contains a summary and conclusions
	\section{Stochastic mortality models}
	\label{s:Stoch}
		In publications on mortality modeling and life forecasting, many stochastic models can be found. The dynamic development of this element of demography took place in 1992 due to Lee and Carter's article \cite{LeeCarter}. The authors proposed a stochastic model for which the parameter estimation methodology refers to the analysis of the main components. Lee and Carter set themselves the goal of determining more precise life expectancy forecasts for the United States. For the research they used data on the number of deaths in the USA from 1900 to 1987. In this methodology, the logarithm of the central death rate $\ln\(m_{x,t}\)$ is explained. The original formulation of the model is
		\[
		\ln\(m_{x,t}\)=\alpha_{x}+\beta_{x}\times 
		\kappa_{t}+\varepsilon_{x,t}\,
		\]
		for $x = 1, . . . , N$ i $t = 1, . . . , T$;
		where
		\begin{itemize}
			\item $m_{x,t}$ -- the central death rate for age $x$ and year $t$,
			\item $\alpha_{x}$ -- represents the general mortality shape across age, equals the average of $\ln\(m_{x,t}\)$ over time $t$,
		\end{itemize}
		\[
		\alpha_{x}=\frac{1}{T}\sum\limits_t {\ln \left( m_{x,t} \right),} 
		\]
		\begin{itemize}
			\item $\kappa_{t}$ -- represents the time trend,
			\item $\beta_{x}$ -- is an age-specific parameter, representing the sensitivity of the log of the mortality rates at age $x$ to the time trend represented by $\kappa_{t}$,
			\item $\varepsilon_{x,t}$ -- the error term, assumed to be i.i.d. with mean 0 and constant variance.
			
		\end{itemize}
		The $ \kappa_{t} $ and $ \beta_{x} $ parameters are not unique. If we put $\tilde{\kappa }_{t}=c\kappa_{t}$ and $\tilde{\beta }_{x}=\beta_{x}/c$ in places $\kappa_{t}$ and $\beta_{x}$ we will get the same values as $\ln\(m_{x,t}\)$. For this reason, two conditions are introduced to ensure the model identification:
		\[
		\sum\limits_t \kappa_{t} =0,\quad\sum\limits_x \beta_{x} =1.
		\]
		The assumptions (confirmed by subsequent studies) regarding the autocorrelation and heteroskedasticity of the random component limited the possibility of using such estimation methods as the classically least squares method (\cite{Alho}, \cite{Brouhns}, \cite{Olivieri}). In addition, the presence of the bilinear term $\beta_{x}\kappa_{t}$ causes some problems in the estimation of parameters. Lee and Carter partially solved these problems using a method that refers to the singular value decomposition (SVD) . This approach requires assumption of homoscedasticity of the random component $\varepsilon_{x,t}$. Research suggests that the variance is not equally distributed in the sample (\cite{Alho}, \cite{Brouhns}). This phenomenon is clearly visible, for example, when comparing the variance in the age groups of 30--50 and 80--100 years. An alternative to the SVD method is the maximum likelihood method. In this estimation method, we assume that the number of deaths is a random variable with the Poisson distribution (\cite{Brillinger}, \cite{Brouhns}), more specifically:
		\[
		D_{x,t}\sim Poisson(\hat{m}_{x,t} E_{x,t})
		\]
		where $\hat{m}_{x,t}=\exp(\hat{\alpha }_{x}+\hat{\beta}_{x}\hat{\kappa }_{t})$, and ${E}_{x,t}$ means the number of exposure-to-risk. The essence of this method is to maximize the logarithm of the likelihood function, which in the case of the Lee-Carter model has the form:
		\[
		\ln L=\sum\limits_x \sum\limits_t 
		\left[D_{x,t}\ln\(m_{x,t}E_{x,t}\)
		-E_{x,t}
		\exp \left(\alpha_{x}+\beta_{x}\kappa_{t}\right)-\ln\(D_{x,t}!\)\right].
		\]
		In the equation is the product of the components $ \beta $ and $\kappa $ which can not be estimated using standard methods. The solution to this problem is the application of the iterative approach. In each iteration step, one of the parameters is estimated, with the remaining values unchanged, in accordance with the general rule:
		\[
		\hat{\theta }^{(v+1)}
		=\hat{\theta }^{(v)}
		-\frac{\partial L^{\(v\)} / \partial \theta } 
		{\partial^{2}L^{\(v\)}/ 
			{\partial \theta^{2}}}\,.
		\]
		For each parameter, the formula has the form:
		\begin{itemize}
			\item $\hat{\alpha }_x^{(v+1)}=\hat{\alpha }_x^{(v)}-\frac{\displaystyle\sum\nolimits_t {(D_{x,t}-\hat{D}_{x,t}^{\left( v \right)})} }{\displaystyle\sum\nolimits_t \hat{D}_{x,t}^{\left( v \right)} }$\,,
			\item $\hat{\kappa }_t^{(v+1)}
			=\hat{\kappa }_t^{(v)}
			-\frac{\displaystyle\sum\nolimits_x {(D_{x,t}
					-\hat{D}_{x,t}^{\left( v \right)})
					\hat{\beta }_{x}^{(v)}} }{\displaystyle\sum\nolimits_x \hat{D}_{x,t}^{\left( v \right)} {(\hat{\beta }_{x}^{\left( v \right)})}^{2}}$\,,
			\item $\hat{\beta }_x^{(v+1)}=\hat{\beta }_x^{(v)}-\frac{\displaystyle\sum\nolimits_t {(D_{x,t}-\hat{D}_{x,t}^{\left( v \right)})\hat{\kappa }_{t}^{(v)}} }{\displaystyle\sum\nolimits_t \hat{D}_{x,t}^{\left( v \right)} {(\hat{\kappa }_{t}^{\left( v \right)})}^{2}}$\,.
		\end{itemize}
	One of the biggest advantages of this model is the ability to make accurate forecasts and determine prediction errors. To get forecasts, you should extrapolate $\kappa_{t} $ - a parameter that describes changes in mortality over time. It can be assumed that $ \kappa_{t} $ is a random walk with a drift, i.e.
	\[
	\kappa_{t}=\kappa_{t-1}+c+\xi_{t}\,,
	\]
	where $\xi_{t}\quad-$ are independent random errors with identical normal distributions 
	$N\left( 0,\sigma^{2} \right)$
	
	The estimation of the parameters $ c $ and $ \sigma^{2} $ can be done as in the publication of N. Li, R. Lee, S. Tuljapurkara \cite{Li}. Estimators of these parameters are given by:
	
	\[
	\hat{c}=(\kappa_{T}-\kappa_{1})/(T-1)
	\]
	and
	\[
	\hat{\sigma^{2}}=1/(T-1)\sum\limits_{t=2}^T {{(\kappa_{t}-\kappa 
			_{t-1}-\hat{c})}^{2}.} 
	\]
	It is possible to extrapolate the $ \kappa $ parameter at the moment T $+$ s:
	\[
	\kappa_{T+s}=\kappa_{T}+\left( \hat{c}+sc\cdot \eta \right)
	s+\hat{\sigma } \sum\limits_{\tau =T+1}^{T+s} {\xi_{\tau },} 
	\]
	where $\eta $ has a normal distribution $N(0,1)$ and $sc\approx \hat{\sigma 
	}/\sqrt {T-1} $ is an estimation error of $c$.

	In the literature there have been many modifications of the Lee-Carter model (\cite{Cairns}, \cite{Plat}, \cite{Renshaw}). 
	The modifications consisted mainly in changing the analytical form of the model either by adding new linear factors or by dividing existing parameters to factors corresponding to characteristic age groups. The first procedure should give even better decomposition of the variance of modeled variable, while the second responded to the postulate often appearing in the literature, that the model should not be built for a full age group, but focus on individual subgroups such as teenagers or the elderly. An interesting modification was proposed in 2006 by Renshaw and Haberman \cite{Renshaw}. They noticed that part of the mortality variance may be the result of a cohort effect (the effect of belonging to the group of people born in the same year).
	The model is given by the following formula:
	\[
	\ln\(m_{x,t}\)= \alpha_{x}+\beta 
	_{x}^{0}\gamma_{t-x}+\beta_{x}^{1}\kappa_{t}+\varepsilon_{x,t}\,.
	\]
	Interpretation of the parameters $\alpha_{x}$, $\beta_{x}^{1}$ i $\kappa_{t}$  is the same as in the Lee-Carter model. The additional parameter $ \gamma_{t-x} $ in the Renshaw-Haberman model describes changes in the mortality as a result of belonging to the cohort of people born in the year $ t-x $. In this case also, there is a problem with parameter identification, which is why the following conditions are added:
	$\sum_x \beta_{x}^{0}=1$, $\sum_x \beta_{x}^{1}=1$
	and $\gamma_{t_{min}-x_{max}}=0$.
	The authors proposed that estimators of unknown parameters can be found using the iterative method:
	\begin{itemize}
		\item $\hat{\gamma }_{z}^{\(v+1\)}
		=\hat{\gamma }_{z}^{\(v\)}+\frac{\displaystyle\sum\nolimits_{z=t-x} {(D_{xt}-\hat{D}_{xt}^{\left( v \right)})\hat{\beta }_{x}^{0^{\(v\)}}} }{\displaystyle\sum\nolimits_{z=t-x} \hat{D}_{xt}^{\(v\)} ({\hat{\beta }_{x}^{0^{\(v\)}})}^{2}}$\,,
		\item $\hat{\beta}_x^{{0}^{\(v+1\)}}=\hat{\beta }_x^{0^{\(v\)}}+\frac{\displaystyle\sum\nolimits_t {\(D_{xt}-\hat{D}_{xt}^{\left(v \right)}\)\hat{\gamma }_{t-x}^{\(v\)}} }{\displaystyle\sum\nolimits_t \hat{D}_{xt}^{\(v\)} ({\hat{\gamma }_{t-x}^{\(v\)})}^{2}}$\,,
		\item $\hat{\beta }_x^{{1}^{\(v+1\)}}
		=\hat{\beta }_x^{1^{\(v\)}}+\frac{\displaystyle\sum\nolimits_t {\(D_{xt}-\hat{D}_{xt}^{\left(v \right)}\)\hat{\kappa }_{t}^{\(v\)}} }{\displaystyle\sum\nolimits_t \hat{D}_{xt}^{\(v\)} ({\hat{\kappa }_{t}^{\(v\)})}^{2}}$\,,
		\item $\hat{\kappa }_t^{\(v+1\)}
		=\hat{\kappa }_t^{\(v\)}
		+\frac{\displaystyle\sum\nolimits_x {(D_{xt}-\hat{D}_{xt}^{\left( v \right)})\hat{\beta }_{x}^{{1}^{(v)}}}}
		{\displaystyle\sum\nolimits_x \hat{D}_{xt}^{(v)} 
			\({\hat{\beta}_x^{1^{\(v\)})}}\)^2
		}$\,.
	\end{itemize}
	The model is extrapolated in the same way as the Lee-Carter model. The authors proposed, in later works, an extension of this
	model based on the assumption that the improvement in mortality should not only be analyzed in terms of calendar years, but also in a cohort terms. Renshaw and Haberman try to model mortality in a cohort terms, not how it is usually done in the age-calendar year \cite{Renshaw}. 
	
	Another modification of the Lee-Carter model is the model proposed by Plata in 2009 \cite{Plat}. The change in this model consists in adding another age-dependent parameter to describe changes in mortality in younger age groups. The new component, which is to capture the specific character of changes in the mortality rate of young people, still allows the use of a model for the entire age range. The model is given by the following formula:
	\[
	\ln \left( m_{x,t} \right)=\alpha_{x}+\kappa_{t}^{1}+\kappa_{t}^{2}\left( 
	\bar{x}-x \right)+\kappa_{t}^{3}{(\bar{x}-x)}^{+}+\gamma_{t-x}+\varepsilon 
	_{x,t}
	\]
	where $(\bar{x}-x)^{+}=\max \left( \bar{x}-x,0 \right)$.
	The model parameters are estimated using iterative methods:
	\begin{itemize}
		\item $\hat{\gamma }_{z}^{(v+1)}=\hat{\gamma }_{z}^{(v)}+\frac{\displaystyle\sum\nolimits_{z=t-x} {(D_{xt}-\hat{D}_{xt}^{\left( v\right)})} }{\displaystyle\sum\nolimits_{z=t-x} \hat{D}_{xt}^{(v)} }$\,,
		\item $\hat{\kappa }_t^{{1}^{(v+1)}}
		=\hat{\kappa }_t^{1^{(v)}}
		+\frac{\displaystyle\sum\nolimits_x {\(D_{xt}-\hat{D}_{xt}^{\left( v \right)}\)} }
		{\displaystyle\sum\nolimits_x \hat{D}_{xt}^{\(v\)} }
		$\,,
		\item $\hat{\kappa }_t^{{2}^{(v+1)}}=\hat{\kappa }_t^{2^{(v)}}+\frac{\displaystyle\sum\nolimits_x {(D_{xt}-\hat{D}_{xt}^{\left( v \right)})(\bar{x}-x)} }{\displaystyle\sum\nolimits_x \hat{D}_{xt}^{\left( v \right)} (\bar{x}-{x)}^{2}}$\,,
		\item $\hat{\kappa }_t^{{3}^{(v+1)}}=\hat{\kappa }_t^{3^{(v)}}+\frac{\displaystyle\sum\nolimits_x {(D_{xt}-\hat{D}_{xt}^{\left( v \right)})(\bar{x}-{x)}^{+}} }{\displaystyle\sum\nolimits_x \hat{D}_{xt}^{\left( v \right)} \{(\bar{x}-{{x)}^{+}\}}^{2}}$\,.
	\end{itemize}
	Researchers are still trying to improve the Lee-Carter model. There is a suggestion to use new classes of time series models to predict values of $ \kappa_{t} $ e.g. Generalized Autoregressive Score (GAS) or Dynamic Conditional Score (DCS) models \cite{Neves}. 
	Some scientists try to use the Lee-Carter approach not only to model the mortality rate, but to model changes in the mortality rate, eg by predicting changes $\ln\(m_{x,t+1}\) - \ln\(m_{x,t}\)$ in the central death rate \cite{Mitchell}. 
	
	\section{Modeling for the number of deaths in a heterogeneous population}
	\label{s:Model}
	
	In the early papers of researchers analyzing mortality and life expectancy (eg Halley) one can find a statement that the populations are heterogeneous due to mortality (\cite{Pitacco}, \cite{Jasiulewicz}).
	In a large group of used models, it is assumed that the number of deaths $ D $ in the population has a Poisson distribution:
	\[
	D\sim Poisson(\lambda ).
	\]
	The $ \lambda $ parameter describes the risk of death in the entire population.
	Research has shown that the assumption of Poisson distribution is biologically and substantively justified for groups of people of similar age \cite{Brouhns}. It is a widely used practice to generalize this assumption to the entire population. However, a detailed analysis of the mortality rate in different age groups indicated the existence of a large variation in variance. This means that mortality modeling methods should take into account heterogeneity of the population.
	
	Let us assume that the random variable $ \Lambda $ with distribution function $ F \left (\lambda \right) $ determines the structure of mortality in the population.
	The problem comes down to the question: What is the distribution of the variable $ D $ if the heterogeneity is described by the distribution function $ F \left (\lambda \right) $?
	In the insurance practice, the gamma distribution is often used to describe the heterogeneity of the client population \cite{Lord}. 
	The conditional distribution of the number of deaths is given by the following formula:
	\[
	p_{d}\left( \lambda \right)=\Pr\left( D_{x,t}=d\vert\Lambda=\lambda\right)=\frac{\lambda^{d}}{d!}e^{-\lambda },
	\quad 
	d=0,1,2,\dots
	\]
	This corresponds to the assumption of Poisson's distribution of deaths in age-homogeneous groups. Using the law of total probability, the distribution of deaths can be obtained:
	\[
	\begin{split}
	p_{d}\left( \lambda \right)
	&=\Pr\left( D_{x,t}=d \right)=\int\limits_0^\infty 
	{p_{d}\left( \lambda \right)dF(\lambda )} =\int\limits_0^\infty {\frac{\lambda 
			^{d}}{d!}e^{-\lambda }dF\left( \lambda \right)} \\
	&=
	\frac{\mathrm{\Gamma }(d+1)}{d!\mathrm{\Gamma }(\alpha )}\left( \frac{\tau 
	}{1+\tau } \right)^{\alpha }\left( \frac{1}{1+\tau } \right)^{d}.
	\end{split}
	\]
	Additionally, you can calculate the expected value and the variance of the number of deaths:
	\[
	\E D_{x,t}=\E\Lambda= \frac{\alpha }{\tau }\,,
	\]
	\[
	\Var D_{x,t}=\Var\Lambda=\frac{\alpha }{\tau^{2}}\,.
	\]
	 Like Brouhns N., Denuit M., Vermunt J.,\cite{Brouhns} estimating the model
	\[
	\ln\(m_{x,t}\)=\alpha_{x}+\beta_{x}\kappa_{t}+\varepsilon_{x,t}
	\]
	should start with writing the following relationship:
	\[
	D_{x,t}=\hat{m}_{x,t} E_{x,t}\,.
	\]
	To calibrate the model with historical data and to ensure that the expected values match, let us assume that: $\alpha =E_{x,t}$ and $\tau = e^{-\(\alpha _{x}+\beta_{x}\kappa_{t}\)}$.
	In the next step we use the method of the maximum likelihood as a method of estimating the parameters of the model.The maximum likelihood function is given as 
	\[
	\begin{split}
	\ln L=const
	&-\sum_x \sum_t {E_{x,t}\exp\left(\alpha_{x}+\beta_{x}\kappa_{t} \right)} \\
	&-\sum_x \sum_t \ln\left( 1+e^{-\(\alpha_{x}+\beta_{x}\kappa_{t}\)} \right) 
	\(E_{x,t}+D_{x,t}\).
	\end{split}
	\]
	Due to the complicated analytical form of the function $\ln L$ (presence of the bilinear term $\beta_{x}\kappa_{t}$) the iterative method was used to estimate unknown parameters $\alpha_{x}\beta_{x}$, $\kappa_{t}$.
	The iterative scheme looks as follows
	\begin{itemize}
		\item $\hat{\alpha }_x^{(v+1)}=\hat{\alpha }_x^{(v)}-\frac{\displaystyle\sum_t \frac{\hat{E}_{x,t}-E_{x,t}}{1+\exp\({-\alpha }_{x}-\beta_{x}\kappa_{t}\)} }{\displaystyle\sum_t \frac{-\hat{E}_{x,t}+E_{x,t}\exp({-\alpha }_{x}-\beta_{x}\kappa_{t})}{({1+\exp\({-\alpha }_{x}-\beta_{x}\kappa_{t}\))}^{2}} }$\,,
		\item $\hat{\kappa }_t^{(v+1)}
		=\hat{\kappa }_t^{(v)}-\frac{\displaystyle\sum\nolimits_x \frac{(\hat{E}_{x,t}-E_{x,t})\hat{\beta }_{x}}{1+\exp({-\alpha }_{x}-\beta_{x}\kappa_{t})} }
		{\displaystyle\sum_x \frac{(-\hat{E}_{x,t}+E_{x,t}\exp \left( {-\alpha }_{x}-\beta_{x}\kappa_{t} \right))\hat{\beta }_{x}^{2}}{({1+\exp({-\alpha }_{x}-\beta_{x}\kappa_{t}))}^{2}} }$\,,
		\item $\hat{\beta }_x^{(v+1)}=\hat{\beta }_x^{(v)}-\frac{\displaystyle\sum\nolimits_t \frac{(\hat{E}_{x,t}-E_{x,t})\hat{\kappa }_{t}}{1+\exp({-\alpha }_{x}-\beta_{x}\kappa_{t})} }{\displaystyle\sum\nolimits_t \frac{(-\hat{E}_{x,t}+E_{x,t}\exp \left( {-\alpha }_{x}-\beta_{x}\kappa_{t} \right))\hat{\kappa }_{t}^{2}}{({1+\exp({-\alpha }_{x}-\beta_{x}\kappa_{t}))}^{2}} }$\,.
	\end{itemize}

	\section{Numerical applications}
		In this section, I am considering the Lee-Carter model describing the rates of mortality for Polish female population. The data are from 1959 to 2009 and was taken from Human Mortality Database. Two cases will be considered: in the first variant, the number of deaths is described by the Poisson distribution and in the second by negative-binomial distribution.
		\begin{figure}[htbp]
			\centering 
			\includegraphics[width=12cm]{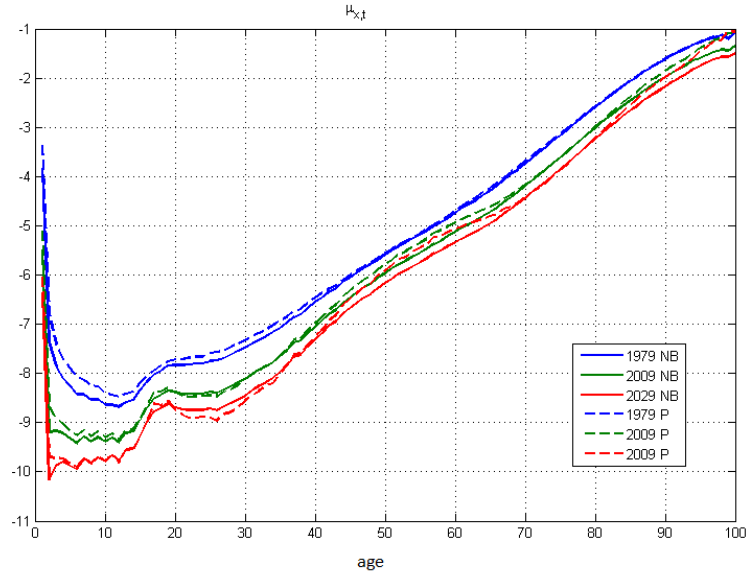}
			\caption{\label{fig1}Logarithms of mortality rates. Number of deaths modeled by Poisson distribution (P), negative-binomial distribution (NB)}
		\end{figure}
		Figure 1 presents the results of estimation for three selected calendar years: 1979, 2009 and 2029 (I have only put three examples to make the figure easier to analyze). The logarithm of the death rate for 2029 is a forecast created in accordance with the method described in section~\ref{s:Stoch}.
		In the first two cases, the mortality rate calculated in the model with the number of deaths described by the negative  binomial distribution is lower or equal to the mortality rate calculated in the model with the number of deaths described by the Poisson distribution. The difference is significant especially in the group of young people under 30. Historical data confirms that in the youngest group the decrease in mortality over the last 50 years was the most significant. Previous studies on UK data also indicated that the Lee-Carter model (if the number of deaths has a Poisson distribution) underestimates the decline in the mortality rate in the youngest group. This may suggest that the Lee-Carter model better informs about significant changes in mortality rates in subpopulations when the negative binomial distribution of deaths is used. For the forecast, similar results can be seen for all cohorts, except for cohorts from 20 to 30 years old. 
		
		The sums of residual squares may indicate a better fit of the model to the data. Results for several selected calendar years are presented below.
		\begin{table}[htbp]
			\caption{\label{tab}Fitting the model - sum of residual squares $\sum{e^2}$}
			\begin{center}
				\begin{tabular}{|l|l|l|l|l|l|}
					\hline
					Model& 1980& 1990& 1995& 2000& 2005 \\
					\hline
					Poisson& 3.40& 2.19& 1.61& 1.91& 5.17 \\
					\hline
					Negative binomial& 1.02& 1.04& 1.29& 1.01& 1.72 \\
					\hline
				\end{tabular}
				\label{tab1}
			\end{center}
		\end{table}
		In all cases the Lee-Carter model with the number of deaths described by negative binomial distribution better adapts to historical data.
	
	\section{Conclusion}
	The research shows that Lee-Carter models can be successfully used in mortality modeling. The maximum likelihood method can be used to estimate structural parameters. However, not only the Poisson distribution, but also other distributions should be used in the modeling the number of deaths.  In the case of heterogeneous populations, the conducted research has shown that good results can be achieved by the use of negative binomial distribution. The fit of the Lee-Carter model in this case is clearly better.
	
	\section*{Acknowledgment}
	This work is supported by National Science Centre Grant Nos. 2014/13/N/HS4/03192

\end{document}